\documentclass[11pt,twoside]{article}

\usepackage{asp2006}
\usepackage{epsf}
\usepackage{psfig}
\usepackage{lscape}

\begin{document}
\title{Mid-Infrared Evidence for Accelerated Evolution in Compact Group Galaxies}   
\author{Lisa May Walker$^1$, Kelsey E. Johnson$^1$, Sarah C. Gallagher$^2$, John E. Hibbard$^3$, Ann E. Hornschemeier$^4$, Jane C. Charlton$^5$, Thomas H. Jarrett$^6$}
\affil{$^1$Dept of Astronomy, University of Virginia, Charlottesville, VA 22904; $^2$Dept of Physics and Astronomy, University of Western Ontario, London, ON N6A 3K7 Canada; $^3$NRAO, Charlottesville, VA 22903; $^4$Laboratory for X-Ray Astrophysics, NASA Goddard Space Flight Center, Greenbelt, MD 20771; $^5$Dept of Astronomy and Astrophysics, Pennsylvania State University, University Park, PA 16802; $^6$\textit{Spitzer} Science Center, California Institute of Technology, Pasadena, CA 91125}

\begin{abstract} 
We find evidence for accelerated evolution in compact group galaxies from the distribution in mid-infrared colorspace of 42 galaxies from 12 Hickson Compact Groups (HCGs) compared to the the distributions of several other samples including the LVL+SINGS galaxies, interacting galaxies, and galaxies from the Coma Cluster. We find that the HCG galaxies are not uniformly distributed in colorspace, as well as quantitative evidence for a gap. Galaxies in the infall region of the Coma cluster also exhibit a non-uniform distribution and a less well defined gap, which may reflect a similarity with the compact group environment. Neither the Coma Center or interacting samples show evidence of a gap, leading us to speculate that the gap is unique to the environment of high galaxy density where gas has not been fully processed or stripped.
\end{abstract}


\section{Samples}
The HCG dataset, taken from \citet{hcgs}, comprises 42 galaxies from 12 groups. The groups contain varying amounts of HI, and span the three classification categories discussed in \citet{hcgs}. The most HI gas-rich groups with $\log{\left( M_{\mathrm{H \, I}} \right)}/\log{\left( M_{\mathrm{dyn}} \right)} \ge 0.9$ are classified as type I , the HI gas-poor groups with $\log{\left( M_{\mathrm{H \, I}} \right)}/\log{\left( M_{\mathrm{dyn}} \right)} < 0.8$ as type III, while type II contains the intermediate groups.

The LVL data, discussed in \citet{lvl} consists of 211 galaxies within 11 Mpc. The SINGS data, as described in \citet{sings}, consist of 71 galaxies. We combined the LVL and SINGS galaxies to create a control sample, referred to as LVL+SINGS. The 35 galaxies from the Spitzer Spirals, Bridges, and Tails Interacting Galaxy Survey (hereafter referred to as the interacting sample) are comprised of otherwise relatively isolated binary galaxy systems, whose members are tidally disturbed \citep{arp}. The Coma sample, discussed in \citet{coma}, is comprised of galaxies from two fields. The first field is located in the center of the cluster, where the galaxy density is very high. The second field is the infall region, located near 0.4 virial radii at the X-ray secondary peak, where the galaxy density is still higher than field density \citep{coma}. Our HCG sample has very few galaxies with luminosities below $\log{\left(L_{4.5}\;[\mathrm{erg/s}]\right)} = 40.9$. Therefore, to compare the HCG galaxies to similar galaxy populations from the other samples we only consider galaxies with luminosities greater than this.

\section{Colorspace}
Galaxies with blue MIR colors have IR SEDs consistent with being dominated by stellar light. Galaxies with ongoing SF tend to have red MIR colors indicative of PAH emission and warm/hot dust. The gap between these colors in HCG colorspace may be due to rapid evolution through the stage during which galaxies have intermediate MIR colors. The dotted lines indicate the range of comparison used for the statistical tests.

\begin{figure}[t]
 \plottwo{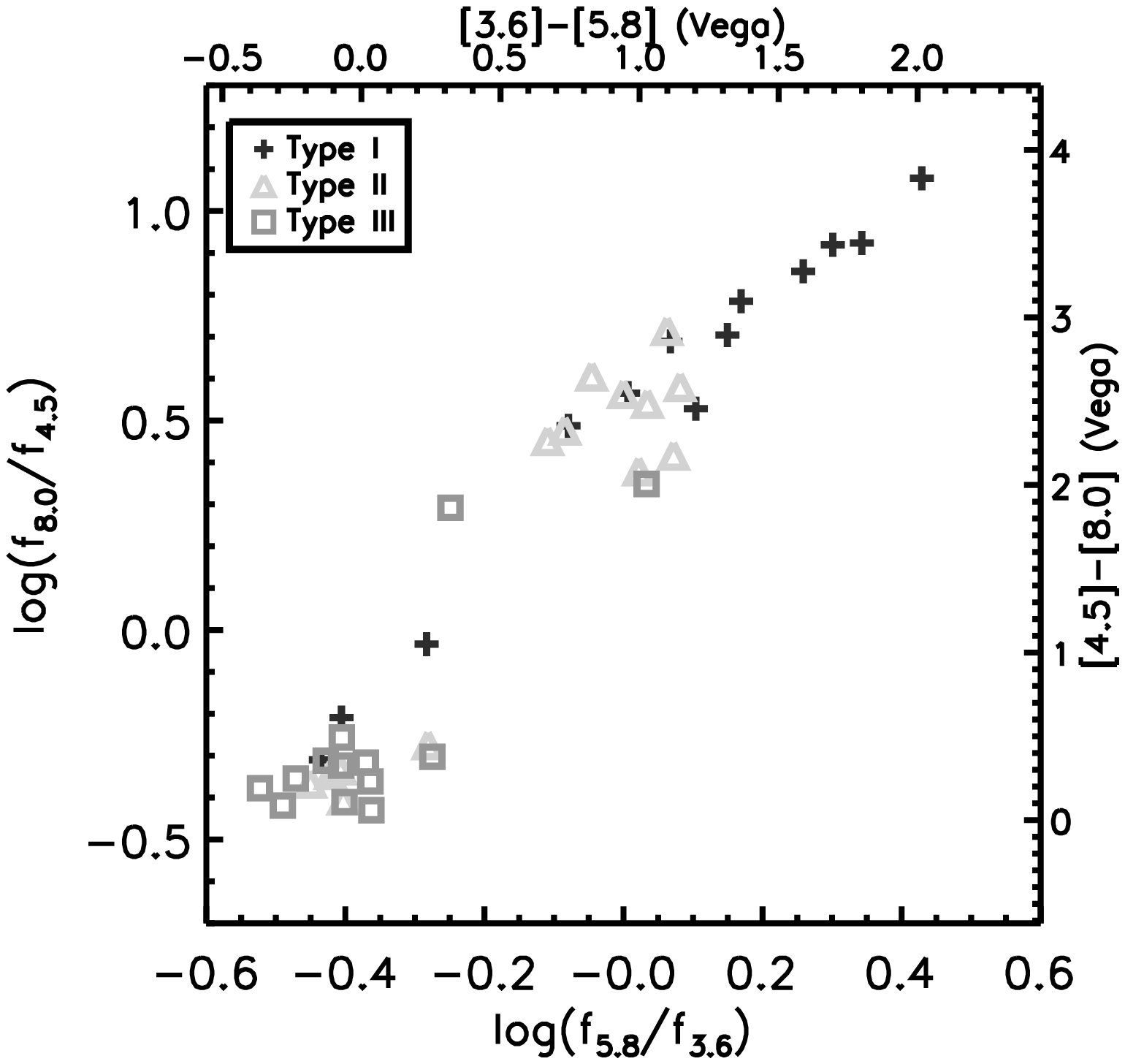}{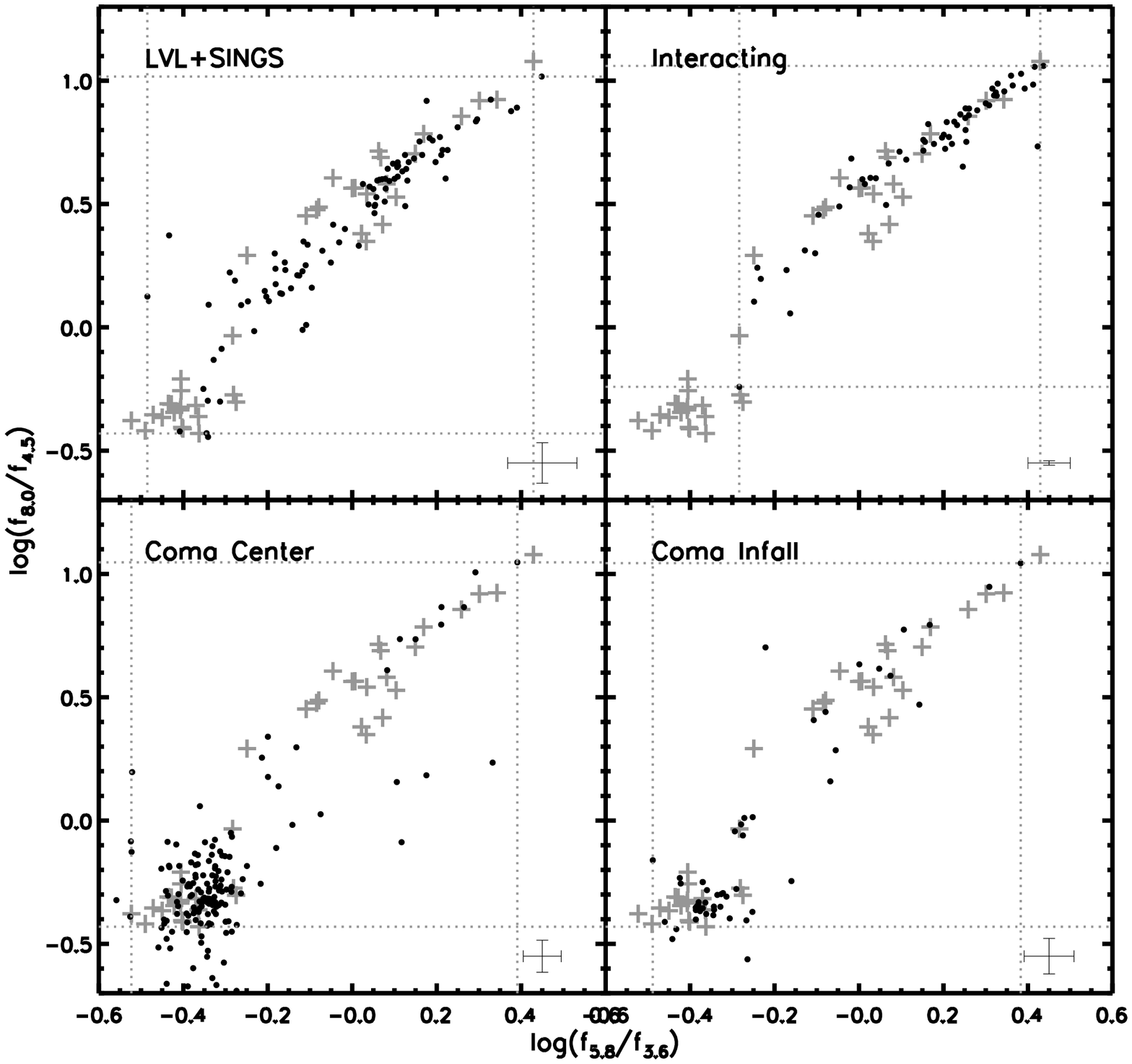}
 \caption{{\itshape Left:\/} Color-color plot of HCG galaxies. The plus signs are galaxies from type I (HI gas-rich) groups, the triangles correspond to galaxies from type II groups, and the squares represent galaxies from type III (HI gas-poor) groups. The lower-left region of the plot contains galaxies whose light is dominated by normal stellar populations, while active galaxies (i.e. star-forming) reside in the upper right.
 {\itshape Right:\/} Color-color plot of comparison samples. The grey plus signs are the HCG galaxies, and the error bars in the bottom right indicate typical errors for each sample. The dashed lines indicate the cropping values, so that the analysis was done only over the range of colorspace occupied by both the HCG galaxies and the comparison sample.\label{colorspace}}
\end{figure}

\section{Statistical Tests}
The KS test comparing the HCG galaxies with a uniform distribution clearly illustrates the gap, manifested as the nearly-horizontal portion of the CDF. The two-distribution KS tests yield $\alpha$ values for the infall Coma sample high enough that we cannot reject the hypothesis that they are drawn from the same parent distribution as the HCGs over the range of comparison.

\begin{figure}[t]
 \plottwo{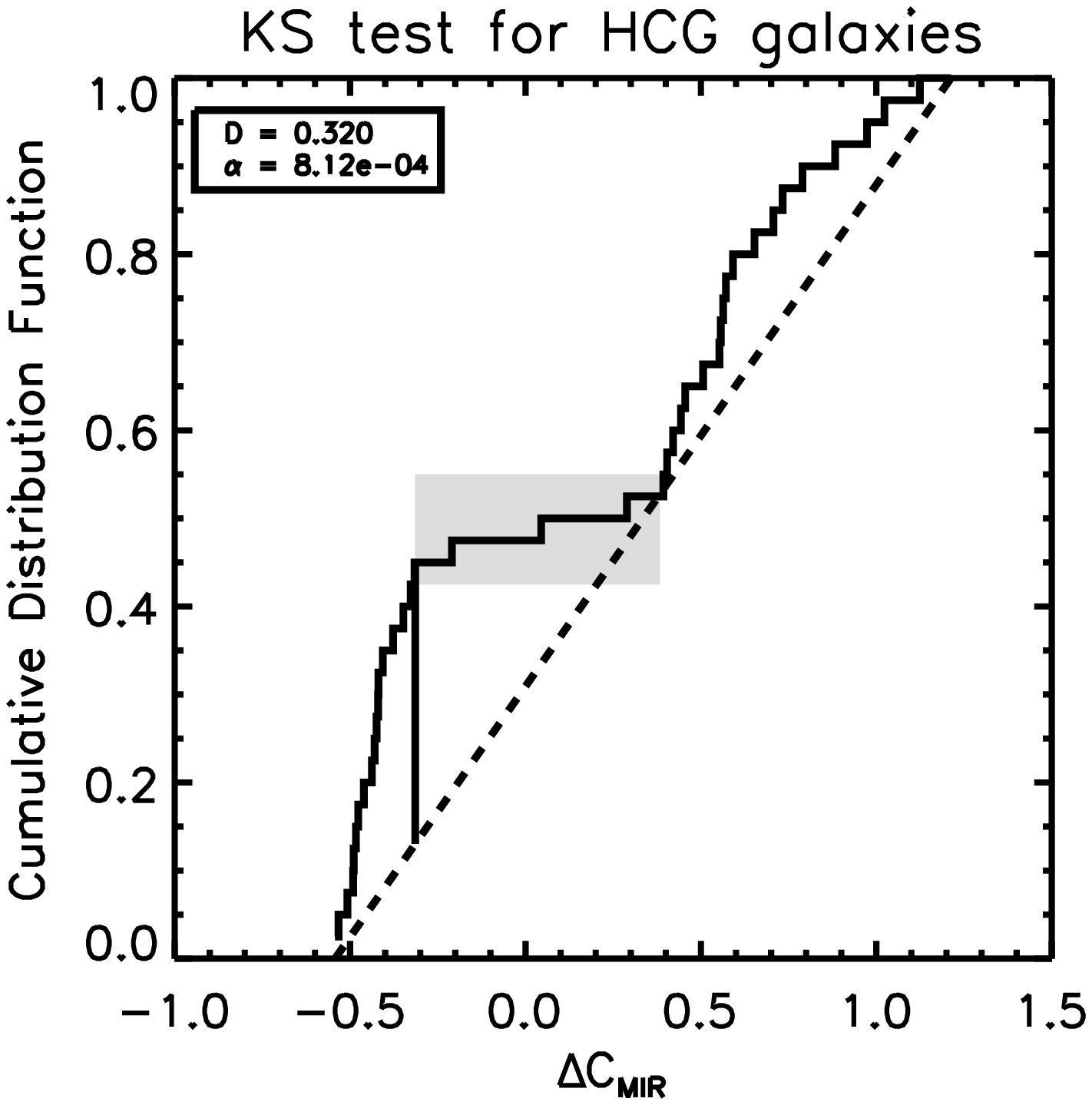}{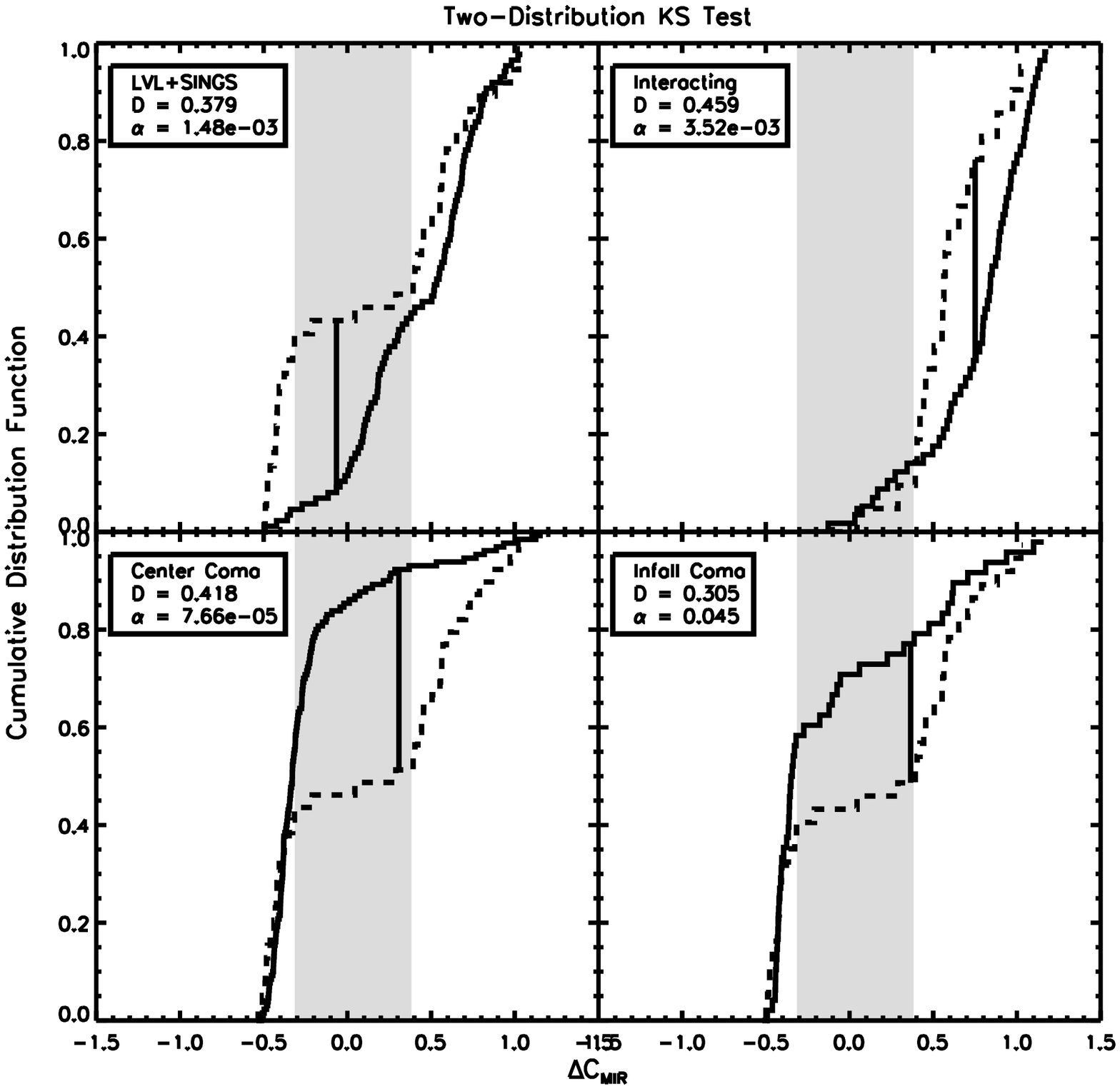}
 \caption{{\itshape Left:\/} KS test for the rotated HCG distribution against a model of uniform distribution. The maximum deviation of the HCG CDF (solid line) from a uniform distribution (dashed line) is $D$, indicated by the vertical line. The nearly flat portion highlighted in grey reveals the gap.
 {\itshape Right:\/} KS test for the rotated comparison samples against the HCG galaxies.\label{ks}}
\end{figure}

\section{Gap Region}
The gap region is defined by HCG galaxies bounding gap at $\Delta_{C_{MIR}}= -0.31$ and $0.38$, indicated by the gray box in Figure \ref{ks}. 
Since the gap represents a deficit of galaxies over this color range, its signature is a flattening in the CDFs. 
We excluded the interacting sample due to insufficient data blueward of gap
The HCG sample has the most pronounced gap, followed by Coma Infall.
\begin{table}[!ht]
\caption{Slopes of the CDFs of all samples over the region of the HCG gap.\label{gapslope}}
\smallskip
\begin{center}
{\small
\begin{tabular}{cc}
\tableline
\noalign{\smallskip}
Sample &	Slope\\
\noalign{\smallskip}
\tableline
\noalign{\smallskip}
HCG		&	0.104	 \\
LVL + SINGS	&	0.808	 \\
Coma (center)	&	0.544	 \\
Coma (infall)	&	0.215	 \\
\noalign{\smallskip}
\tableline
\end{tabular}
}
\end{center}
\end{table}

\section{CMDs}
The morphology of the CMDs reflects the criteria used to define each sample. For example, both Coma samples have a significant concentration of galaxies with blue MIR colors, characteristic of galaxies with little to no SF, while the samples biased toward actively star forming systems show a scattering of galaxies with red MIR colors. The CMD of the HCGs is a composite of these two types, but has a deficit of galaxies with intermediate MIR colors $\rightarrow$ a ÒtransitionalÓ phase between no PAHs and strong PAH features.

\begin{figure}[t]
 \plottwo{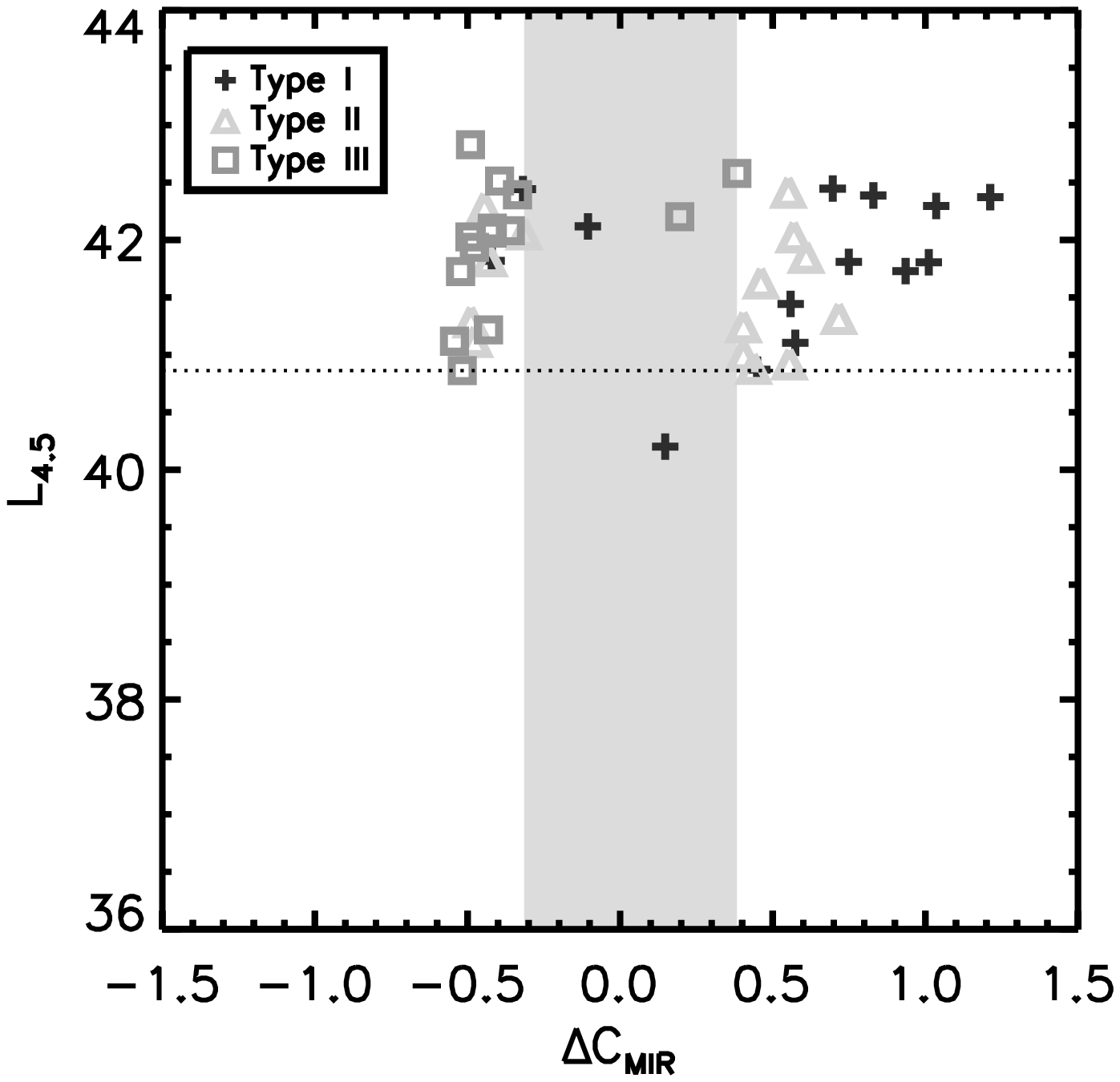}{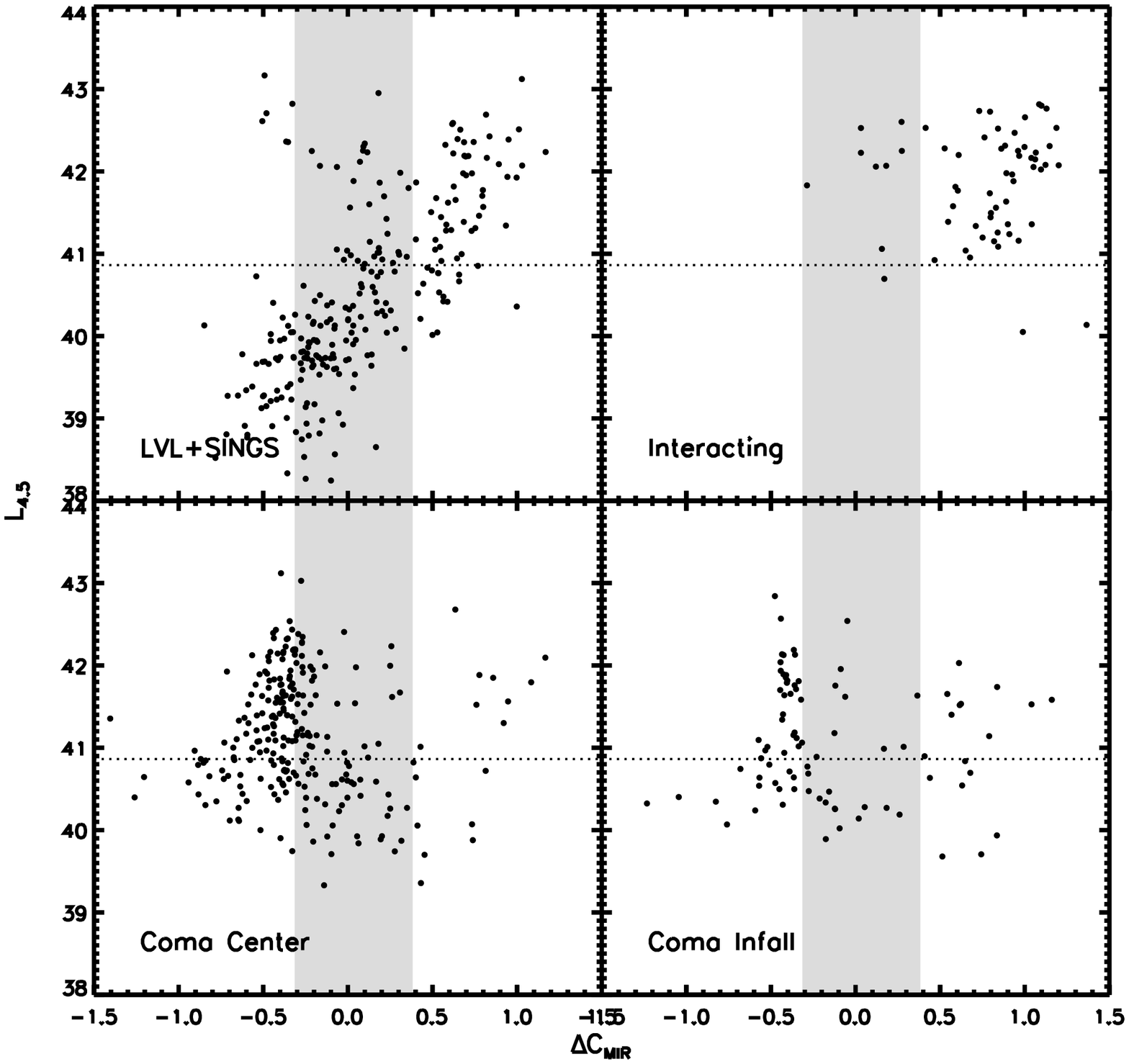}
 \caption{{\itshape Left:\/} CMD for the HCG galaxies. The dotted line indicates the minimum luminosity required for inclusion. The gap is indicated by the grey stripe. It is noteworthy that there does not seem to be a color-luminosity correlation.
 {\itshape Right:\/} CMD for the comparison samples. The criteria used to define each sample is apparent in the morphology of the CMDs.\label{cmd}}
\end{figure}

\section{Discussion}
We suggest that the distribution in MIR colorspace, and particular the occurance of a gap in the color distribution, likely reflect different star formation states, with galaxies ranging from active/efficient (red MIR colors) to passive/inefficient (blue MIR colors). The distribution in colorspace suggests that galaxies in the HCG environment move quickly from active to passive states, in a manner similar to, but more dramatic than, the environment on the outskirts of clusters. Galaxies from both more dense (Coma center) and less dense (e.g. LVL+SINGS) environments have a continuous distribution of systems between active and passive states.

The presence of the gap in the MIR colorspace distribution of the HCGs combined with the fact that the gap is not present in less dense environments indicates that this type of local environment significantly influences galaxy properties. In order to understand the processes that affect galaxy evolution, we need to understand how gas is processed in the interstellar medium and intragroup medium in these environments. Compact groups are clearly an important part of understanding galaxy assembly and hierarchical structure formation.


\end{document}